\UseRawInputEncoding 
\documentclass[superscriptaddress,aps,prl,reprint,twocolumn,amsmath,amssymb,showpacs,]{revtex4-2}
\usepackage{graphicx}  
\usepackage{mathptmx}
\usepackage{epstopdf}
\usepackage{dcolumn}   
\usepackage{bm}        
\usepackage{amssymb}   
\usepackage{amsmath}   
\usepackage{amsthm}
\usepackage{chemarrow}
\usepackage{color}
\usepackage{mathrsfs}
\usepackage{float}
\usepackage{bbold}
\usepackage{bbm}
\usepackage[a4paper,colorlinks=true,
linkcolor=blue,citecolor=blue,
pdfauthor={ },
pdftitle={ },
pdfsubject={ },
pdfkeywords={ }]{hyperref}

\theoremstyle{plain}
\newtheorem*{theorem*}{Theorem}

\begin{document}


\title{Search for exotic spin-dependent interactions with a spin-based amplifier}

\date{\today}

\author{Haowen Su}
\email[]{These authors contributed equally to this work}
\affiliation{
Hefei National Laboratory for Physical Sciences at the Microscale and Department of Modern Physics, University of Science and Technology of China, Hefei, Anhui 230026, China}
\affiliation{
CAS Key Laboratory of Microscale Magnetic Resonance, University of Science and Technology of China, Hefei, Anhui 230026, China}
\affiliation{
Synergetic Innovation Center of Quantum Information and Quantum Physics, University of Science and Technology of China, Hefei, Anhui 230026, China}

\author{Yuanhong Wang}
\email[]{These authors contributed equally to this work}
\affiliation{
Hefei National Laboratory for Physical Sciences at the Microscale and Department of Modern Physics, University of Science and Technology of China, Hefei, Anhui 230026, China}
\affiliation{
CAS Key Laboratory of Microscale Magnetic Resonance, University of Science and Technology of China, Hefei, Anhui 230026, China}
\affiliation{
Synergetic Innovation Center of Quantum Information and Quantum Physics, University of Science and Technology of China, Hefei, Anhui 230026, China}

\author{Min Jiang}
\email[]{dxjm@ustc.edu.cn}
\affiliation{
Hefei National Laboratory for Physical Sciences at the Microscale and Department of Modern Physics, University of Science and Technology of China, Hefei, Anhui 230026, China}
\affiliation{
CAS Key Laboratory of Microscale Magnetic Resonance, University of Science and Technology of China, Hefei, Anhui 230026, China}
\affiliation{
Synergetic Innovation Center of Quantum Information and Quantum Physics, University of Science and Technology of China, Hefei, Anhui 230026, China}

\author{\mbox{Wei Ji}}
\affiliation{Dapartment of Physics, Tsinghua University, Beijing, 100084, China}

\author{Pavel~Fadeev}
\affiliation{Helmholtz-Institut, GSI Helmholtzzentrum f{\"u}r Schwerionenforschung, Mainz 55128, Germany}
\affiliation{Johannes Gutenberg University, Mainz 55128, Germany}

\author{Dongdong Hu}
\affiliation{
State Key Laboratory of Particle Detection and Electronics, University of Science and Technology of China, Hefei, Anhui 230026, China}

\author{Xinhua Peng}
\email[]{xhpeng@ustc.edu.cn}
\affiliation{
Hefei National Laboratory for Physical Sciences at the Microscale and Department of Modern Physics, University of Science and Technology of China, Hefei, Anhui 230026, China}
\affiliation{
CAS Key Laboratory of Microscale Magnetic Resonance, University of Science and Technology of China, Hefei, Anhui 230026, China}
\affiliation{
Synergetic Innovation Center of Quantum Information and Quantum Physics, University of Science and Technology of China, Hefei, Anhui 230026, China}

\author{Dmitry Budker}
\affiliation{Helmholtz-Institut, GSI Helmholtzzentrum f{\"u}r Schwerionenforschung, Mainz 55128, Germany}
\affiliation{Johannes Gutenberg University, Mainz 55128, Germany}
\affiliation{Department of Physics, University of California, Berkeley, CA 94720-7300, USA}

\begin{abstract}
Development of new techniques to search for particles beyond the standard model is crucial for understanding the ultraviolet completion of particle physics.
Several hypothetical particles are predicted to mediate exotic spin-dependent interactions between particles of the standard model that may be accessible to laboratory experiments.
However, laboratory searches are mostly conducted for static spin-dependent interactions, with only a few experiments so far addressing spin- and velocity-dependent interactions.
Here, we demonstrate a search for exotic spin- and velocity-dependent interactions with a spin-based amplifier.
Our technique makes use of hyperpolarized nuclear spins as a pre-amplifier to enhance the effect of pseudo-magnetic field produced by exotic interactions by an amplification factor of $> 100$.
Using such a spin-based amplifier, we establish constraints on the spin- and velocity-dependent interactions between polarized and unpolarized nucleons in the force range of 0.03-100~m.
Our limits represent at least two orders of magnitude improvement compared to previous experiments.
The established technique can be further extended to investigate other exotic spin-dependent interactions.
\end{abstract}

\maketitle

\noindent
\textbf{INTRODUCTION}

\noindent
Numerous theories extending beyond the standard model of particle physics predict the existence of new bosons that could mediate long-range interactions between two objects of the standard model~\cite{ramsey1979tensor,wilczek1978problem,peccei1977cp,weinberg1978new,fadeev2019revisiting}.
Such theories are related with the spontaneous breaking of continuous symmetries, yielding massless or light (pseudo) Nambu-Goldstone bosons,
such as axions~\cite{peccei1977cp,wilczek1978problem}, dark photons~\cite{an2015direct}, paraphotons~\cite{dobrescu2005massless}, familons~\cite{ammar2001search}, and majorons~\cite{dearborn1986astrophysical}.
The exchange of such particles results in exotic spin-dependent interactions that may be accessible to laboratory experiments~\cite{demille2017probing,safronova2018search}.
Various experiments have been conducted to search for exotic spin-dependent interactions,
making use of
the torsion balance~\cite{terrano2015short,ding2020constraints},
trapped ions~\cite{wineland1991search,kotler2015constraints},
geoelectrons~\cite{hunter2013using,hunter2014using},
spin-exchange relaxation-free magnetometers~\cite{kim2019experimental,ji2018new},
comagnetometers~\cite{almasi2020new,lee2018improved},
nitrogen-vacancy diamond~\cite{rong2018searching,rong2018constraints,jiao2020experimental},
and other high-sensitivity techniques~\cite{bulatowicz2013laboratory,tullney2013constraints,ficek2017constraints,ficek2018constraints,ni1999search,kimball2010constraints}.
Recently, the approach based on nuclear magnetic resonance (NMR)~\cite{arvanitaki2014resonantly,chu2020search,aggarwal2020characterization} has been proposed to search for exotic interactions and could substantially improve current experimental limits set by astrophysics.

The exotic interactions mediated by light bosons were introduced by Moody and Wilczek~\cite{moody1984new},  extended by Dobrescu and Mocioiu with the inclusion of the terms dependent on the relative velocity between the two interacting particles~\cite{dobrescu2006spin}, and revised in Ref.~\cite{fadeev2019revisiting}. In the latter work~\cite{fadeev2019revisiting}, the potentials were sorted by types of coupling (scalar, vector, etc), contact interactions were included, and coordinate-space representation were used.
Sorting by spin-momentum form, there are fifteen possible exotic interactions between ordinary
particles that contain static spin-dependent operators, velocity-dependent operators or combinations of these.
Some of them may break the charge, parity and time-reversal symmetries
or their combinations~\cite{dobrescu2006spin,safronova2018search}; they were introduced to understand the symmetries of charge conjugation and parity in quantum chromodynamics (QCD)~\cite{peccei1977cp}.
Many experiments have been performed to search for static spin-dependent interactions~\cite{lee2018improved,kotler2015constraints,rong2018searching,bulatowicz2013laboratory,rong2018constraints,tullney2013constraints,almasi2020new,hunter2013using,wineland1991search,terrano2015short,bulatowicz2013laboratory,ni1999search},
while the velocity-dependent interactions have been studied less extensively~\cite{jiao2020experimental,ji2018new,kim2019experimental,ficek2018constraints,kimball2010constraints}.
Following the notation in Refs.~\cite{dobrescu2006spin,leslie2014prospects}, the spin- and velocity-dependent interactions to be
studied here are
\begin{flalign}
    \label{E1}
    &V_{4+5}=-f_{4+5}\frac{\hbar^2}{8 \pi m c}[\hat{\sigma} \cdot  (\bm{v} \times \hat{r} ) ] 
    \left( \frac{1}{\lambda r}+\frac{1}{r^2} \right) e^{-r/\lambda},& \\
    \label{E2}
    &V_{12+13}=f_{12+13}\frac{\hbar}{8 \pi}(\hat{\sigma} \cdot \bm{v})\left (\frac{1}{r}\right)e^{-r/\lambda},&
\end{flalign}
where $f_{4+5}$, $f_{12+13}$ are dimensionless coupling constant, $c$ is the speed of light in vacuum, $\hat{\sigma}$ is the spin vector and $m$ is the mass of the polarized fermion, $\bm{v}$ is the relative velocity between two interacting fermions, $\hat{r}$ is the unit vector in the direction between them, and $\lambda \!= \!\hbar(m_b c)^{-1}$ is the force range (or the boson Compton wavelength) with $m_b$ being the light boson mass.
In particular, the search for $V_{12+13}$ could provide a new source for parity symmetry violation~\cite{dobrescu2006spin,safronova2018search}.
Careful investigations on those interactions may give clues for understanding fundamental physical questions like matter-antimatter asymmetry of the Universe. 


Here, we demonstrate a search for the exotic spin- and velocity-dependent interactions with a spin-based amplifier~\cite{jiang2021search}.
The technique takes advantage of the resonant coupling between the rotation frequency of an unpolarized nucleon mass and a spin-based amplifier with a matching spin-precession frequency.
The signal from the pseudo-magnetic field produced by the exchange of light bosons can be amplified by at least two orders of magnitude.
Using such a sensor,
we obtain new constraints on the spin- and velocity-dependent interactions ($V_{4+5}$ and $V_{12+13}$) with at least two orders of magnitude improvement over previous works~\cite{piegsa2012limits,haddock2018search,yan2013new,yan2015searching}.
We emphasize the difference between this work and other resonant searches.
Similar NMR-based resonant techniques searching for exotic spin-dependent interactions have been proposed~\cite{arvanitaki2014resonantly,chu2020search,aggarwal2020characterization}.
These references consider the situation where the exotic pseudo-magnetic fields are measured from their proximity with atomic and SQUID magnetometers;
in this case, it is experimentally difficult to prepare high nuclear-spin polarization and maintain readout sensitivity.
In contrast, our work uses a different scheme in which polarized nucleons and the detector are spatially overlapping in the same vapor cell,
offering a significant advantage: nuclear spin signals can be enhanced due to large Fermi-contact enhancement factor (on the order of 600),
measured $in~situ$ with an atomic magnetometer.

\begin{figure}[t]  
	\makeatletter
	\def\@captype{figure}
	\makeatother
	\includegraphics[scale=1.05]{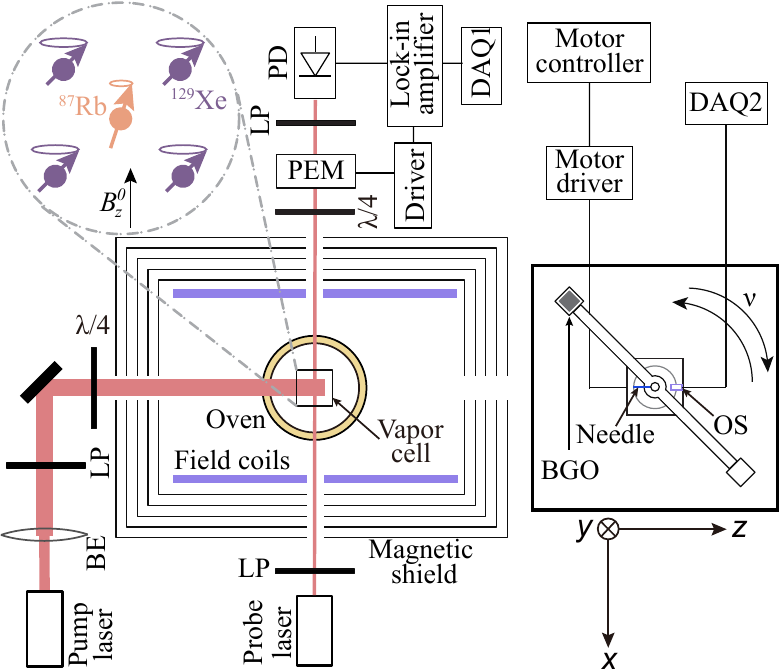}
	\caption{\textbf{Experimental setup}. The $^{87}$Rb magnetometer uses a 0.5~$\textrm{cm}^3$ cubic cell consisting of 5~torr isotopically riched $^{129}$Xe, 250~torr $\textrm{N}_2$ as buffer gas and a droplet of $^{87}$Rb. The vapor cell is placed inside a five-layer cylindrical $\mu$-metal shield to reduce the ambient magnetic field. A bias field $B_z^0\hat{z}$ is applied along $z$ to tune the $^{129}$Xe Larmor frequency to $\nu_0 \approx$4.997~Hz. The $^{87}$Rb spins are polarized by optical pumping with 795~nm D1 light. $^{87}$Rb-$^{129}$Xe spin-exchange collisions polarize $^{129}$Xe spins to $\sim$30$\%$. The $x$ component of $^{87}$Rb spins is measured via optical rotation of a linearly polarized probe beam, which is blue-detuned 110~GHz to $^{87}$Rb D2 transition at 780~nm. The right inset shows the configuration of a bismuth germanate insulator [$\textrm{Bi}_4 \textrm{Ge}_3 \textrm{O}_{12}$ (BGO)] mass and a motor. A single BGO mass at the end of an aluminum rod rotates with frequency $\nu \approx 4.997$~Hz to generate the spin- and velocity-dependent interactions. BE, beam expander; LP, linear polarizer; $\lambda /4$, quarter-wave plate; PD, photodiode; PEM, photoelastic modulator; DAQ, data acquisition; OS, optoelectronic switch.}
	\label{setup}
\end{figure}

~\

\noindent
\textbf{RESULTS}

\noindent
\textbf{Principle}.~We consider exotic pseudo-magnetic fields produced by exotic spin- and velocity-dependent interactions.
The interactions induce energy shift of $^{129}$Xe spins \mbox{$-\! \bm{\mu}_{\textrm{Xe}} \!\cdot\! \textbf{B}_{j}^{\textrm{exo}}\! =\!V_{j}$}~\cite{ni1999search}, where $\bm{\mu}_{\textrm{Xe}}$ is the magnetic moment of $^{129}$Xe spin, $V_{j}$ represents the potential that we measure ($V_{4+5}$ or $V_{12+13}$) and $\textbf{B}_{j}^{\textrm{exo}}$ is the pseudo-magnetic field produced by the exotic interaction $V_{j}$.
We assume that the coupling strengths between different polarized nucleons (neutron and proton) and the unpolarized nucleons of the mass source are the same.

We now consider the resonant response of the spin-based amplifier to the pseudo-magnetic field.
The key element of the spin-based amplifier is the spatially overlapping ensembles of spin-polarized $^{87}$Rb and $^{129}$Xe, as shown in Fig.~\ref{setup}. 
The exotic interaction can generate an oscillating pseudo-magnetic field $\textbf{B}_j^{\textrm{exo}}$ on $^{129}$Xe spins that slightly tilts $^{129}$Xe spins and induces an oscillating $^{129}$Xe transverse magnetization that is read out by $^{87}$Rb spins. 
The spin dynamics can be described by the coupled Bloch equations~\cite{jiang2019floquet,jiang2021search,ji2018new},
\begin{flalign}
    \label{E3}
    &\frac{\partial \textbf{P}^{e}}{\partial t}=\frac{\gamma_{e}}{Q} (B^{0}_z\hat{z}+\beta {M}^{n}_0\textbf{P}^{n})\times\textbf{P}^{e} + \frac{P^{e}_{0} \hat{z} -\textbf{P}^{e} }{T_e Q},&\\
    \label{E4}
    &\frac{\partial \textbf{P}^{n}}{\partial t}\!=\!\gamma_{n}({B}^{0}_z\hat{z}\!+\!\textbf{B}_{j}^{\textrm{exo}}\!+\!\beta {M}^{e}_0\textbf{P}^{e})\!\times\!\textbf{P}^{n}\!+\!\frac{P^{n}_{0}\hat{z}\!-\!\textbf{P}^{n} }{\{T_{2n},T_{2n},T_{1n}\}},&
\end{flalign} 
where $\textbf{P}^{e}$ ($\textbf{P}^{n}$) is the polarization of $^{87}$Rb electron ($^{129}$Xe nucleus); $\gamma_e$ ($\gamma_n$) is the gyromagnetic radio of the $^{87}$Rb electron ($^{129}$Xe nucleus); $Q$ is the electron slowing-down factor originated from hyperfine interaction and spin-exchange collisions; $B^0_z\hat{z}$ is the applied bias field; $M^{e}_0$ ($M^{n}_0$) is the maximum magnetization of $^{87}$Rb electron ($^{129}$Xe nucleus) associated with full spin polarizations; $P^e_0$ ($P^n_0$) is the equilibrium polarization of the $^{87}$Rb electron ($^{129}$Xe nucleus); $T_e$ is the common relaxation time of $^{87}$Rb electron spins; and $T_{1n}$ ($T_{2n}$) is the longitudinal (transverse) relaxation time of $^{129}$Xe spins. 
The Fermi-contact interaction between $^{87}$Rb and $^{129}$Xe pairs introduces an effective magnetic field $\textbf{B}^{e,n}_{\textrm{eff}}=\beta M_0^{e,n}\textbf{P}^{e,n}$, where $\beta = 8\pi \kappa_0 /3$~\cite{walker1997spin,jiang2019floquet,jiang2021search}.
The effective field generated by $^{129}$Xe spins is read out $in$ $situ$ with the $^{87}$Rb magnetometer.

We first consider the resonant response of the spin-based amplifier to a single-frequency component, for example, $\textbf{B}_{\textrm{ac}}^{\textrm{exo}}=B_{\textrm{ac}}^{\textrm{exo}} \cos(2 \pi \nu t)\hat{y}$ of the pseudo-magnetic field $\textbf{B}_j^{\textrm{exo}}$. 
In this situation, the bias field $B_z^0\hat{z}$ is tuned to satisfy the resonant condition $\nu=\nu_0=\gamma_n B_z^0 /(2 \pi)$. 
Due to negligible $\textbf{B}^{e}_{\textrm{eff}}$ and relatively strong bias field $B^0_z\hat{z}$, $^{129}$Xe spins independently evolve without the influence of $^{87}$Rb spins. 
For small transverse excitations of $^{129}$Xe spins, we can derive the steady-state solution of $^{129}$Xe transverse polarization from Eq.~(\ref{E4}). 
A detailed derivation can be found in Supplementary Materials (section S1) and Ref.~\cite{jiang2021search}.
The corresponding transverse magnetization generates a measurable oscillating effective magnetic field on $^{87}$Rb magnetometer. 
As a result, the amplitude of the effective field is 
\begin{equation}
    \label{E5}
    |\textbf{B}^n_{\textrm{eff}}| = \dfrac{4 \pi}{3}  \kappa_0 M^{n}_0 P^{n}_{0} \gamma_{n} T_{2n} |\textbf{B}_{\textrm{ac}}^{\textrm{exo}}|.
\end{equation}
We define a factor $\eta=\dfrac{4 \pi}{3}  \kappa_0 M^{n}_0 P^{n}_{0} \gamma_{n} T_{2n}$ that represents the amplification of the signal from the pseudo-magnetic field. This shows that the signal from the pseudo-magnetic field can be pre-amplified by the hyperpolarized long-lived $^{129}$Xe spins. 
Specifically, due to large Fermi-contact enhancement factor $\kappa_0 \approx 540$ for $^{129}$Xe-$^{87}$Rb system, $\eta$ is estimated to be over $100$~\cite{jiang2021search}.
For $^{3}$He-K system, for which the spin-coherence time is much longer ($\sim 1000$~s)~\cite{almasi2020new,lee2018improved}, $\eta$ can reach $10^4$. 

~\

\begin{figure}[t]  
	\makeatletter
	\def\@captype{figure}
	\makeatother
	\includegraphics[scale=0.95]{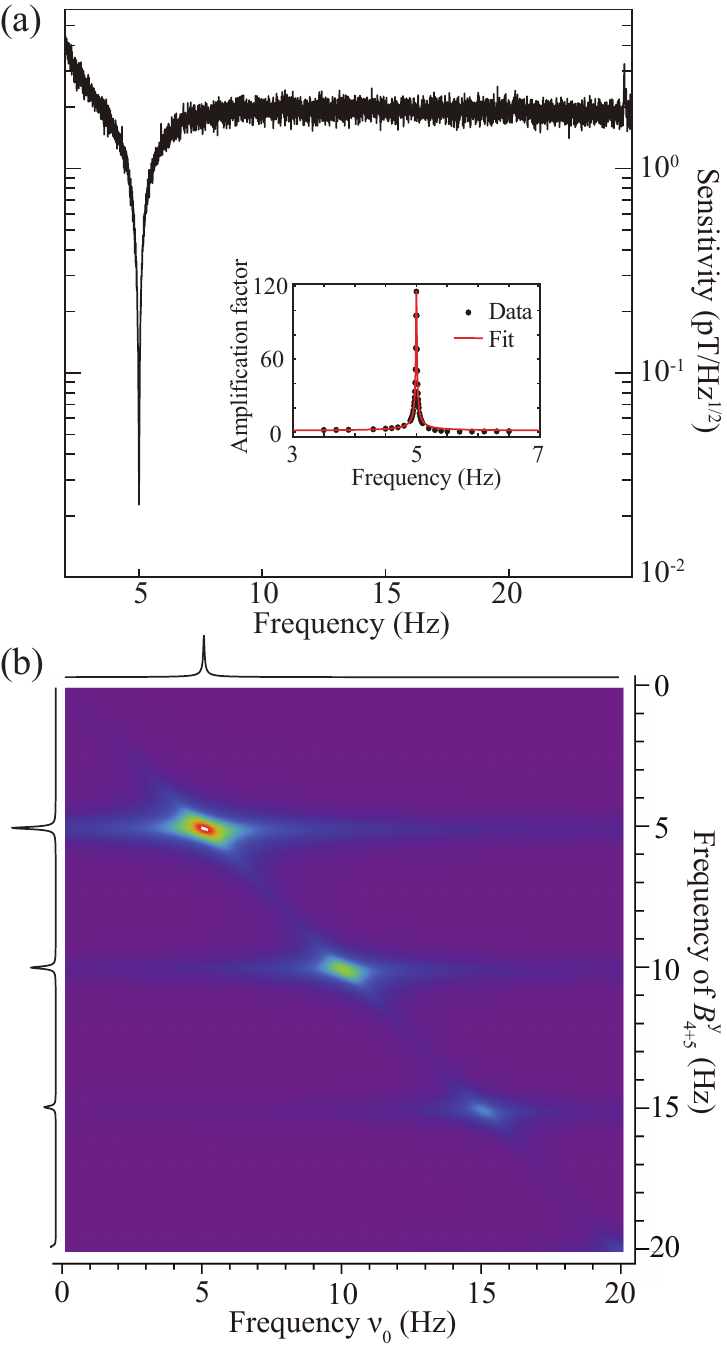}
	\caption{\textbf{Demonstration of the spin-based amplifier and basic principle of exotic interaction searches}. (a) Sensitivity of the spin-amplifier-based magnetometer. The inset exhibits the frequency dependence of the amplification factor. The experimental data (black dots) are obtained by applying a calibration field along $y$. As demonstrated in Supplementary Materials (section S1), the red solid line is a theoretical fit of the data with $A/\sqrt{(\nu-\nu_0)^2+(\Lambda/2)^2}$~\cite{jiang2021search}, where the full width at half maximum (FWHM) is $\sqrt{3} \Lambda \approx 13$~mHz. The enhanced sensitivity of the spin-amplifier-based magnetometer reaches $\approx$~22~fT/$\sqrt{\textrm{Hz}}$ at 4.997~Hz. (b) Three dimensional diagram of the enhanced signals from $B_{4+5}^y$ with the spin-based amplifier. The $x$ axis represents the frequency dependence of the spin-based amplifier [see (a), inset]. The $y$ axis represents the decomposition of the pseudo-magnetic field $B_{4+5}^y$ generated by the single rotating BGO crystal. By scanning the resonant frequency $\nu_0$ of the spin-based amplifier, there are three peaks of enhanced signals at harmonics of $B_{4+5}^y$ with matched frequency $\nu_0=\nu,2\nu,3\nu$. The signal of the pseudo-magnetic field reaches maximum colored with white at 4.997~Hz.}
	\label{sensitivity}
\end{figure}

\noindent
\textbf{Experimental setup}.~Experiments are performed using a setup similar to that of Refs.~\cite{jiang2019floquet,jiang2021search}, depicted in Fig.~\ref{setup}.
We experimentally calibrate the amplification factor $\eta$ and the corresponding enhanced magnetic-field sensitivity of $^{87}$Rb magnetometer.
For example, the bias field $B_z^0$ is set as 423~nT, corresponding to $^{129}$Xe Larmor frequency $\nu_0 \approx 4.997$~Hz. 
An oscillating magnetic field of 13.0~pT is applied along $y$ to simulate the single-frequency pseudo-magnetic field $B_{\textrm{ac}}^{\textrm{exo}} \cos(2 \pi \nu t)\hat{y}$.
By scanning the oscillation frequency $\nu$ near the resonance, the maximum value on resonance is determined as the amplification factor.
As shown in Fig.~\ref{sensitivity}(a), the signal amplitude is well described by a single-pole band-pass filter model (see inset)~\cite{jiang2021search} (see section S1), yielding the full width at half maximum (FWHM) of 13~mHz.
The maximum $\eta \approx 116$ is achieved at $\nu \approx \nu_0 \approx 4.997 $~Hz. 
By taking the response of the spin-based amplifier into account, the magnetic sensitivity of $^{87}$Rb magnetometer is enhanced to $\approx$~22~fT/$\sqrt{\textrm{Hz}}$, whereas the off-resonance sensitivity of $^{87}$Rb magnetometer is only about 2~pT/$\sqrt{\textrm{Hz}}$. 
Moreover, the enhanced sensitivity is far beyond that of the state-of-the-art magnetometers demonstrated with nuclear spins~\cite{gross2016dynamic,wu2018nuclear}, which are usually limited to a few picotesla sensitivity.
Therefore, an atomic magnetometer enhanced with a spin-based amplifier is well suited for resonantly searching for exotic spin- and velocity-dependent interactions.

The pseudo-magnetic field, for example, $\textbf{B}_j^{\textrm{exo}}=\textbf{B}_{4+5}^{\textrm{exo}}$ is generated by the configuration (see Materials and Methods) in Fig.~\ref{setup}.
For a source of unpolarized nucleons, we use a single 112.34-g bismuth germanate insulator [$\textrm{Bi}_4 \textrm{Ge}_3 \textrm{O}_{12}$ (BGO)] crystal with a high number density of nucleons~\cite{kim2019experimental}.
Driven by a servo motor, the single BGO crystal connected to a 48.76-cm aluminum rod rotates with frequency $\nu \approx 4.997$~Hz in the $xz$ plane. 
The rotation frequency and the phase of the pseudo-magnetic field can be measured through the triggered optoelectronic pulses.
The center of the aluminum rod is located 58.32~cm away from the center of the $^{129}$Xe vapor cell.
The rotating BGO crystal generates a pseudo-magnetic field $\textbf{B}_{4+5}^{\textrm{exo}}\!=\!B_{4+5}^y\hat{y}$ along $y$.
As demonstrated in Supplementary Materials (section S2), the field $B_{4+5}^y\hat{y}$ can be decomposed into
\begin{equation}
   B_{4+5}^y = \sum B^{(N)}_{\textrm{ac}} \cos(2 \pi N \nu t),
\end{equation}
where $ N \nu$ is the multiple frequency and $B_{\textrm{ac}}^{(N)}$ is the corresponding field strength.
For example, we consider the case of $\lambda=1.0$~m and $f_{4+5}=1$ for $V_{4+5}$ generated by the BGO crystal [see Eq.~(\ref{E1})].
Based on our numerical simulation (see Materials and Methods),
the ratios of the field strengths at harmonic frequencies are $\!B_{\textrm{ac}}^{(1)}\!:\!B_{\textrm{ac}}^{(2)}\!:\!B_{\textrm{ac}}^{(3)}\! \approx \!5.1\!:\!2.9\!:1.4\!$, as shown in Fig.~\ref{sensitivity}(b) (the $y$ axis).
Accordingly, we choose the dominant ($N=1$) first harmonic $B_{\textrm{ac}}^{(1)} \cos(2 \pi \nu t)\hat{y}$ at 4.997~Hz to be measured.
To this end, we set the operation frequency of the spin-based amplifier at $\nu_0 \approx 4.997$~Hz.
In this situation, due to the relatively narrow bandwidth of the spin-based amplifier, only the signal at $4.997$~Hz can be considerably amplified and the other harmonics are negligible.

\begin{figure}[t]  
	\makeatletter
	\def\@captype{figure}
	\makeatother
	\includegraphics[scale=1.05]{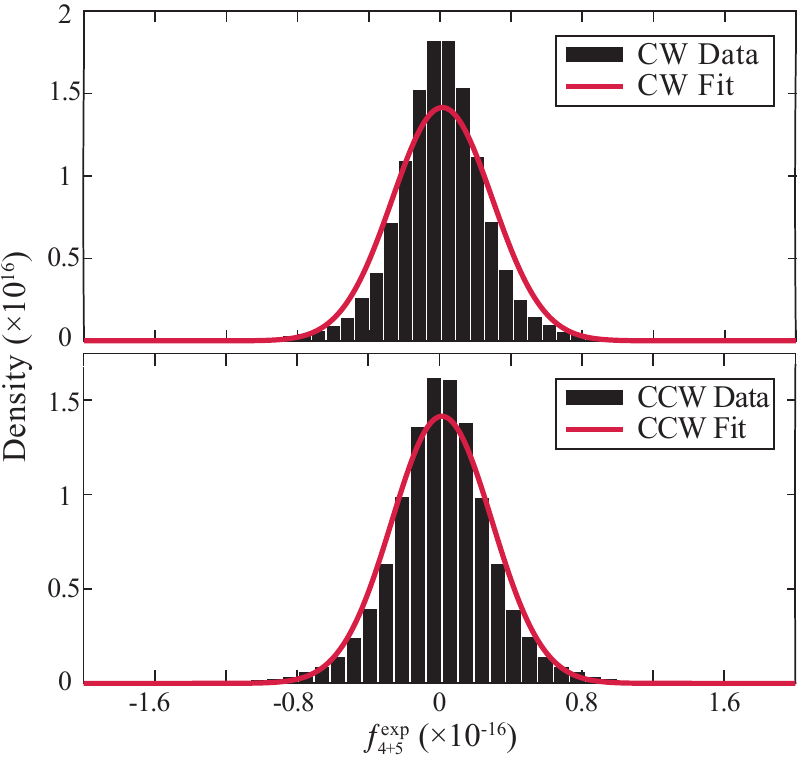}
	\caption{\textbf{Histograms of the potential experimental coupling strength} $f_{4+5}^\textrm{exp}$. Distribution of the experimental coupling strength $\langle f_{4+5} \rangle _{\textrm{CW}}^\textrm{exp}$ for clockwise rotation and $\langle f_{4+5} \rangle _{\textrm{CCW}}^\textrm{exp}$ for counterclockwise rotation.
	The red solid line is a fit to a Gaussian distribution. CW, clockwise; CCW, counterclockwise.}
	\label{cw}
\end{figure}

Considerable effort is made to reduce undesirable noise, such as vibrations, common-mode magnetic noise, electronic cross-talk, etc, before search experiments.
First, a single BGO crystal is used instead of two.
For symmetrically placed two BGO crystals, the dominant signal of the pseudo-magnetic field is at $2\nu$,
at which there is a noise peak due to the air-vibration effect generated by the rotating aluminum rod.
Moreover, an enclosure for the spin-based amplifier setup is used to reduce the air-vibration effect.
We note that the aluminum rod is also a source of unpolarized nucleons, which generates the pseudo-magnetic field consisting of even harmonics at $2N\nu$.
The details are presented in Supplementary Materials (section S2).
Nevertheless, the potential pseudo-magnetic field from the aluminum rod can still be neglected, because the sensitive frequency of the spin-based amplifier is set at $\nu \approx 4.997$~Hz.
Second, the single BGO crystal is rotated clockwise for 1~hour and then counterclockwise for 1~hour, in turn.
As a result, the common-mode magnetic noise can be eliminated by reversing the sign of the pseudo-magnetic field.
Finally, the setup of the spin-based amplifier and the mechanical rotation system are separated.
To suppress the electronic cross-talk caused by the motor and its control system, all equipment related to the setup of the spin-based amplifier is powered from a different circuit to eliminate all connections between them.

~\

\noindent
\textbf{New constraints on $V_{4+5}$ and $V_{12+13}$}.~We perform the searches for $V_{4+5}$, $V_{12+13}$, collecting data for 10~hours. 
Extracting the weak signal with a known carrier frequency from noisy environment is crucial in obtaining the signal from the pseudo-magnetic field $B_{\textrm{ac}}^{(1)} \cos(2 \pi \nu t)\hat{y}$.
We use a ``lock-in'' analysis scheme, similar to the similarity analysis method described in Ref.~\cite{ji2018new}.
The signal can be extracted through a reference signal with the same oscillation frequency. 
Accordingly, the normalized signal of the simulated exotic field at $\lambda =1.0$~m is set as the reference signal $\cos{(2 \pi \nu t+\phi)}$, where $\phi$ is the phase delay between the pseudo-magnetic field and the signal of the $^{87}$Rb magnetometer. 
$\phi$ is calibrated by experiments (see Table I).
Lastly, the experimental coupling strength $f_{4+5}^{\textrm{exp}}$ of one period $T$ is 
\begin{equation}
    f_{4+5}^{\textrm{exp}}= \frac{1}{\alpha B^{(1)}_{\textrm{ac}}} \frac{\int^{T}_0 \cos{(2 \pi \nu t+\phi)} S(t) dt}{\int^{T}_0 \cos^2{(2 \pi \nu t+\phi)} dt},
    \label{E7}
\end{equation}
where $\alpha$ is the calibration factor and $S(t)$ is the experimental signal.
Here, $B^{(1)}_{\textrm{ac}}$ is the field strength simulated with our experimental parameters with the assumption of $\lambda=1.0$~m and $f_{4+5}=1$.
In our experiment, the bound on $f^{\textrm{exp}}_{4+5}$ does not depend on the assumed $\lambda=1.0$~m and $f_{4+5}=1$, i.e., any assumption on these parameters would lead to the same experimental bound.

Based on Eq.~(\ref{E7}), possible values of the potential experimental coupling strength $f_{4+5}^\textrm{exp}$ are calculated for every period $T$.  
Histograms of experimentally measured coupling strength $f_{4+5}^\textrm{exp}$ are collected for 10~hours with clockwise and counterclockwise BGO rotations, as shown in Fig.~\ref{cw}.
Respective fit with Gaussian distribution to each histogram gives $\langle f_{4+5} \rangle _{\textrm{CW}}^\textrm{exp}$ for clockwise cycles and $\langle f_{4+5} \rangle _{\textrm{CCW}}^\textrm{exp}$ for counterclockwise cycles. 
After averaging over all rotating circles, the coupling strength is estimated to be $(2.79 \pm 0.96_{\textrm{stat}}) \times 10^{-19}$.

\begin{table}[t]
\newcommand{\tabincell}[2]{\begin{tabular}{@{}#1@{}}#2\end{tabular}}
\begin{ruledtabular}
\caption {~~~Summary of systematic errors. The corrections to $f_{4+5}^{\textrm{exp}}$ at $\lambda =1.0$~m are listed.} 
\label{tab:my_label}
\renewcommand{\arraystretch}{1.2}
\begin{tabular}{l c c}   
Parameter & Value & $\Delta f^{\textrm{exp}}_{4+5}\left( \times 10^{-19}\right)$ 
  \\
\hline
Mass of BGO (g)& $112.34\pm0.02$ & $\mp 0.001$  \\
Position of pivot $x$ (mm) & $6.0\pm0.3$ & $<0.001$  \\
Position of pivot $y$ (mm) & $3.4\pm0.5$ & $\pm0.001$  \\ 
Position of pivot $z$ (mm) & $583.2\pm1.1$ & $\pm0.012$  \\ 
Length of aluminum rod (mm)& $487.6\pm0.7$ & $\mp 0.013$ \\
Calib. const. $\alpha$ (V/nT) & $6.36^{+0.05}_{-0.93}$ & ${}^{-0.020}_{+0.477}$\\
Phase delay $\phi$ (deg) & $68.0\pm6.0$ & ${}^{+0.100}_{-0.131}$\\[0.2cm]
Final $f^{\textrm{exp}}_{4+5} ( \times 10^{-19})$ & $2.79$
 & $\pm 0.96 \ (\text{stat}) $\\

$(\lambda=1.0 \ \textrm{m})$ &  & $\pm 0.50
\footnote{The origin of coordinates was at the center of the vapor cell.}
$ 
\end{tabular} 
\end{ruledtabular}
\end{table}

Table I summarizes the systematic errors at $\lambda = 1.0$~m in our experiment. 
Due to the narrow bandwidth of the spin-based amplifier, the fluctuation of the rotation frequency is the main systematic error for our experiment, which results in the deviation of the amplification factor and corresponding calibration factor $\alpha$. 
The rotation frequency 4.997$\pm0.004$~Hz results in 14.6$\%$ deviation in calibration factor. 
The overall systematic uncertainty is derived by combining all the systematic errors in quadrature. 
Accordingly, we quote the final total coupling strength $f_{4+5}$ as \mbox{$(2.79 \pm 0.96_{\textrm{stat}} \pm 0.50_{\textrm{syst}}) \times 10^{-19}$}.
Similarly, $f_{12+13}$ is estimated to be \mbox {$(1.64 \pm 0.57_{\textrm{stat}} \pm 0.27_{\textrm{syst}}) \times 10^{-34}$}.

\begin{figure}[t]  
	\makeatletter
	\def\@captype{figure}
	\makeatother
	\includegraphics[scale=1.0]{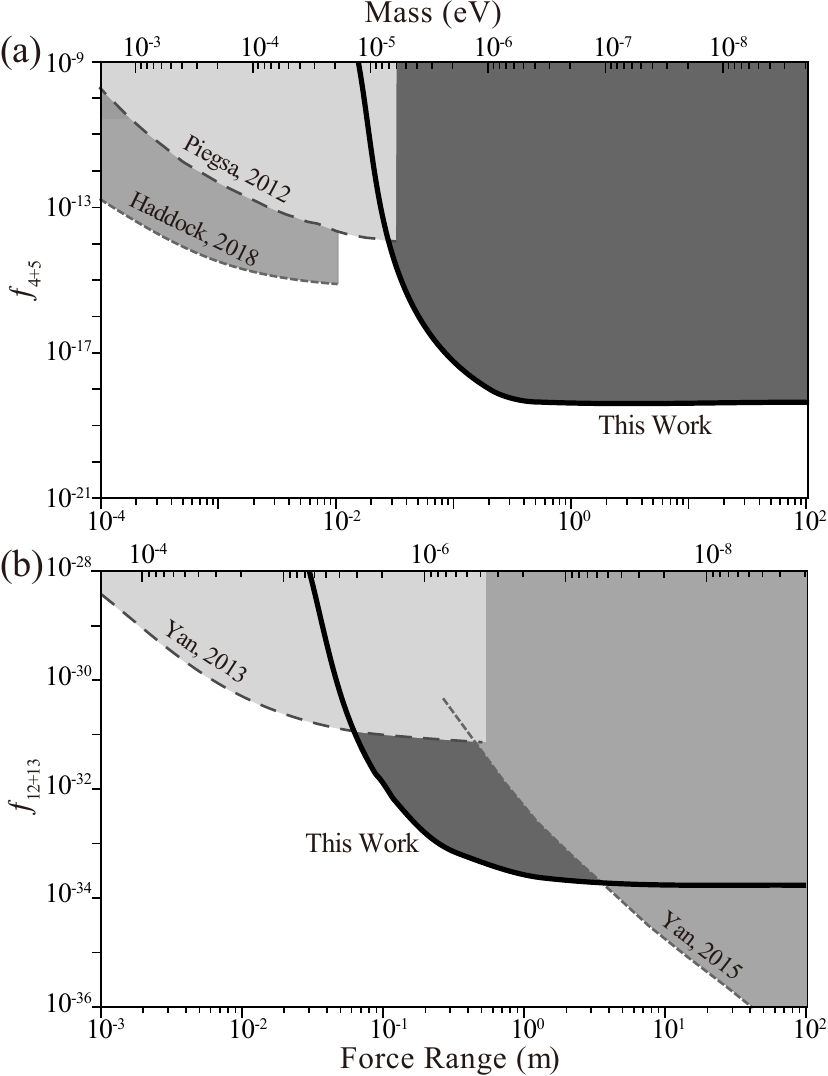}
	\caption{\textbf{Constraints (95$\%$ C.L.) on $f_{4+5}$ and $f_{12+13}$ for the neutron}. In (a), the dashed lines represent bounds of $f_{4+5}$ from Refs.~\cite{piegsa2012limits,haddock2018search}. Our work (solid line) conducts a new laboratory search in the force range from 0.04~m to 100~m. In (b), the dashed lines is from Refs.~\cite{yan2013new,yan2015searching}. The solid line is the constraint of $f_{12+13}$ established by our work, which set the most stringent constraints in the force range from 0.06~m to 4~m. }
	\label{constraint}
\end{figure}

Figure~\ref{constraint} shows the constraints on $V_{4+5}$ and $V_{12+13}$ set by this work. 
Dark areas correspond to excluded values with 95$\%$ confidence level.
The previous constraints of $f_{4+5}$ were established with a cold neutron reflectometer~\cite{piegsa2012limits} ($\lambda <0.04$~m) and a slow neutron polarimeter~\cite{haddock2018search} ($\lambda < 0.01$~m).
In contrast, our bound opens up new experimental searches for the force range from 0.04~m to 100~m, as shown in Fig.~\ref{constraint}(a). 
In Fig.~\ref{constraint}(b), our work constrains $f_{12+13}$ in the force range from 0.03~m to 100~m.
Recent works placed constraints with a cold neutron beam~\cite{yan2013new} ($\lambda<0.06$~m) and polarized $^{3}$He~\cite{yan2015searching} 
($\lambda>4$~m). 
Comparing with them, our work sets the most stringent constraints in the force range from 0.06~m to 4~m, at 0.25~m reaching $1.34\times 10^{-33}$ (95$\%$ confidence level), improving over previous laboratory limits by at least 2 orders of magnitude.

Finally, we link the coupling strength $f_{4+5}$ and $f_{12+13}$ to the appropriate property of the new boson, the product of coupling constants,
\begin{align}
    \label{E8}
    &f_{4+5}= -\frac{1}{2} \frac{g_A^n g_A^n}{\hbar c} -\frac{3}{2} \frac{g_V^n g_V^n}{\hbar c} ,\\
    \label{E9}
    &f_{12+13}= 4 \frac{g_A^n g_V^n}{\hbar c} ,
\end{align} 
where $g_A^n$ is the axial-vector interaction coupling constant and $g_V^n$ is the vector interaction coupling constant for neutrons. 
These couplings originate from an effective Lagrangian of certain new spin-1 bosons, such as the $Z'$ boson~\cite{fadeev2019revisiting,dobrescu2006spin},
\begin{equation} 
 \label{Lag2}
\mathcal{L}_{Z'} = Z'_{\mu} \sum_\psi \bar{\psi}\gamma^{\mu}\left(g_V^\psi + \gamma_{5}  g_A^\psi  \right) \psi  \, ,
\end{equation}
where $\psi$ is the fermion field and $\gamma$ are Dirac matrices.
Through the constraints on $f_{4+5}$ and $f_{12+13}$ established in our experiments, equations~(\ref{E8}) and (\ref{E9}) may be helpful to constrain the interaction coupling constant $g_A^n$ and $g_V^n$.
These investigations are out of the scope of this work and will be discussed in more depth elsewhere.


~\

\noindent
\textbf{Discussion}.~We have reported new constraints,
based on a spin-based amplifier, on exotic spin- and velocity-dependent interactions.
The present experiment opens up new force ranges of searches for $V_{4+5}$ interaction and significantly improves previous limits on $V_{12+13}$ with at least two orders of magnitude.
A further improvement of the experimental sensitivity to spin-dependent interactions can be anticipated.
Increasing the rotor speed and BGO mass could immediately improve the search sensitivity by a factor of 100. 
Further, the application of spin-based amplifier scheme to $^3$He noble gas, which has much longer spin coherence time and larger gyromagnetic ratio than those of $^{129}$Xe, allows one to achieve an amplification factor of $10^4$~\cite{jiang2021search}.
The sensitivity to $V_{4+5}$ and $V_{12+13}$ can be accordingly improved by several orders of magnitude over currently achievable limits. 
In our experiment, a single detector is used but it could be extended to a network of synchronized sensors.
Using such a network around the source mass can enable an improvement of the sensitivity with the inverse square of sensor number and allow to distinguish the potential exotic interaction signal from spurious noise.

This technique can also be used to investigate other exotic spin-dependent interactions.
In particular, our technique is well suited to search for exotic spin-spin-velocity interactions~\cite{ji2018new,hunter2014using,kimball2010constraints},
because they may generate resonantly oscillating pseudo-magnetic fields on spin-based amplifiers.
These exotic interactions are meditated by new bosons,
such as massive spin-1 boson $Z'$ or massless boson $\gamma'$.
Careful investigations on these potentials may unlock some of the deepest puzzles in fundamental physics, such as the strong-$CP$ problem~\cite{peccei1977cp}. 
Therefore, there is growing interest in theories and experiments exploring the frontiers beyond the standard model~\cite{fadeev2019revisiting,hunter2013using,kimball2010constraints,almasi2020new}.
Combining our current $^{129}$Xe spin-based amplifier and recently developed SmCo$_5$ spin sources~\cite{ji2018new,almasi2020new}, the search sensitivities to the spin-spin-velocity-dependent interactions could be improved by several orders of magnitude.

~\

\noindent
\textbf{MATERIALS AND METHODS}

\noindent
\textbf{Simulation of the pseudo-magnetic fields}.
The exotic spin-dependent interactions $V_{4+5}$ and $V_{12+13}$ are generated by a single cube-shaped BGO with $25.0\times25.0\times25.0$~mm$^3$ total volume.
The 112.34-g BGO crystal is concealed into a box at one end of a symmetrical aluminum rod. 
The center of the 48.76-cm aluminum rod is located 58.32~cm away from the center of the $^{129}$Xe vapor cell in alignment along $z$.
The center of the aluminum rod is connected to a servo motor fixed on a stable aluminum platform, through a 31.25-cm titanium cylindrical rod. 
Driven by the motor, the BGO cystal and the aluminum rod rotate with frequency $\nu \approx 4.997$~Hz in $xz$ plane.
The rotation frequency and the phase of the pseudo-magnetic fields can be measured through the triggered optoelectronic pulses.

We simulate the pseudo-magnetic fields $\textbf{B}_{4+5}^{\textrm{exo}}$ and $\textbf{B}_{12+13}^{\textrm{exo}}$ generated by the single rotating BGO crystal. 
The pseudo-magnetic field $\textbf{B}_{4+5}^{\textrm{exo}}$ can be obtained by integrating over the whole volume of the BGO crystal,
\begin{equation}
\begin{split}
     \textbf{B}_{4+5}^{\textrm{exo}}\!=\!f_{4+5}\! \frac{\hbar^2}{8 \pi m c |\bm{\mu}_{\textrm{Xe}}|}\!\iiint \!\rho(\bm{r})\!(\bm{v} \times \hat{r})\!\left(\frac{1}{\lambda r}\!+\!\frac{1}{r^2}\right)\!e^{-r/\lambda}d \bm{r},
\end{split}
\end{equation}
where $\rho(\bm{r})$ is the BGO crystal's nucleon density at location $\bm{r}$ and $m$ is the mass of a neutron.
As demonstrated in Supplementary Materials (section S2), by numerical simulation with the assumption of $\lambda=1.0$~m and $f_{4+5}=10^{-18}$, we show that the pseudo-magnetic field $\textbf{B}_{4+5}^{\textrm{exo}}=B^y_{4+5}\hat{y}$ is not a pure trigonometric function.
The field contains harmonic frequencies at $\nu,2\nu,3\nu,$....
The ratios of the field strengths at harmonic frequencies are $\!B_{\textrm{ac}}^{(1)}\!:\!B_{\textrm{ac}}^{(2)}\!:\!B_{\textrm{ac}}^{(3)}\! \approx 5.1:2.9:1.4$.

The pseudo-magnetic field $\textbf{B}_{12+13}^{\textrm{exo}}$ can be derived by the same method demonstrated above,
\begin{equation}
      \textbf{B}_{12+13}^{\textrm{exo}}=- f_{12+13}  \frac{\hbar}{8 \pi  |\bm{\mu}_{\textrm{Xe}}|} \iiint \rho(\hat{r})(\bm{v} )\left(\frac{1}{r}\right)e^{-r/\lambda}d\bm{r}.
\end{equation}
In contrast to $\textbf{B}_{4+5}^{\textrm{exo}}$, the pseudo-magnetic field $\textbf{B}_{12+13}^{\textrm{exo}}=B_{12+13}^x \hat{x}+B_{12+13}^z \hat{z}$ contains two components along $x$ and $z$.
We note that the spin-based amplifier is insensitive to the oscillating field along $z$, because the oscillating field along $z$ can not induce $^{129}$Xe transverse magnetization. 
Therefore, only $x$ component of the pseudo-magnetic field $B_{12+13}^x\hat{x}$ should be considered. 
Based on our numerical simulation with the assumption of $\lambda=1.0$~m and $f_{12+13}=10^{-33}$,
the ratios of the field strengths at harmonic frequencies are  $\!B_{\textrm{ac}}^{(1)}\!:\!B_{\textrm{ac}}^{(2)}\!:\!B_{\textrm{ac}}^{(3)}\! \approx 5.5:1.7:0.5$, which is demonstrated in Supplementary Materials (section S2).

~\

\noindent
\textbf{SUPPLEMENTARY MATERIALS}

\noindent
Section 1. The response of spin-based amplifier to the exotic spin- and velocity-dependent interactions.\\
Section 2. Numerical simulations of the exotic pseudo-magnetic fields.

\noindent
Figure S1. Experimental setup.\\
Figure S2. Simulations of $V_{4+5}$ produced by a BGO crystal.\\
Figure S3. Simulations of $V_{12+13}$ produced by a BGO crystal.\\
Figure S4. Simulations of $V_{4+5}$ and $V_{12+13}$ produced by an aluminum rod.

\bibliographystyle{naturemag}
\bibliography{mainrefs}

~\

\noindent
\textbf{Acknowledgments}: We thank Yannis K. Semertzidis, \mbox{Xing Rong}, \mbox{Dongok Kim}, YoungGeun Kim, and Yun Chang Shin for valuable discussions.
\textbf{Funding}: This work of H.W.S., Y.H.W., M.J., X.H.P. was supported by National Key Research and Development Program of China (grant no. 2018YFA0306600), National Natural Science Foundation of China (grants nos. 11661161018, 11927811, 12004371), Anhui Initiative in Quantum Information Technologies (grant no. AHY050000),
USTC Research Funds of the Double First-Class Initiative (grant no. YD3540002002).
The work of D.B. and P.F. was supported by the Cluster of Excellence PRISMA+ funded by the German Research Foundation (DFG) within the German Excellence Strategy (Project ID 39083149), by the European Research Council (ERC) under the European Union Horizon 2020 research and innovation program (project Dark-OST, grant agreement No 695405), by the DFG Reinhart Koselleck project.
\textbf{Author contributions}:
M.J., X.H.P., D.B. proposed the experimental concept, devised the experimental protocols, and wrote the manuscript.
H.W.S. and Y.H.W. designed and performed experiments, analyzed the data, and wrote the manuscript.
W.J. and P.F. analyzed the data, added theoretical discussion and edited the manuscript.
D.D.H. assisted in designing the rotating setup.
All authors contributed with discussions and checking the manuscript.
\textbf{Competing interests}: The authors declare that they have no competing interests.
\textbf{Data and materials availability}: All data needed to evaluate the conclusions in the paper are present in the paper and/or the Supplementary Materials.
Additional data related to this paper may be requested to the corresponding author.

\end{document}


\title{Supplementary materials for: \\``Search for exotic spin-dependent interactions with a spin-based amplifier"}

\date{\today}

\author{Haowen Su}
\email[]{These authors contributed equally to this work}
\affiliation{
Hefei National Laboratory for Physical Sciences at the Microscale and Department of Modern Physics, University of Science and Technology of China, Hefei, Anhui 230026, China}
\affiliation{
CAS Key Laboratory of Microscale Magnetic Resonance, University of Science and Technology of China, Hefei, Anhui 230026, China}
\affiliation{
Synergetic Innovation Center of Quantum Information and Quantum Physics, University of Science and Technology of China, Hefei, Anhui 230026, China}

\author{Yuanhong Wang}
\email[]{These authors contributed equally to this work}
\affiliation{
Hefei National Laboratory for Physical Sciences at the Microscale and Department of Modern Physics, University of Science and Technology of China, Hefei, Anhui 230026, China}
\affiliation{
CAS Key Laboratory of Microscale Magnetic Resonance, University of Science and Technology of China, Hefei, Anhui 230026, China}
\affiliation{
Synergetic Innovation Center of Quantum Information and Quantum Physics, University of Science and Technology of China, Hefei, Anhui 230026, China}

\author{Min Jiang}
\email[]{dxjm@ustc.edu.cn}
\affiliation{
Hefei National Laboratory for Physical Sciences at the Microscale and Department of Modern Physics, University of Science and Technology of China, Hefei, Anhui 230026, China}
\affiliation{
CAS Key Laboratory of Microscale Magnetic Resonance, University of Science and Technology of China, Hefei, Anhui 230026, China}
\affiliation{
Synergetic Innovation Center of Quantum Information and Quantum Physics, University of Science and Technology of China, Hefei, Anhui 230026, China}

\author{\mbox{Wei Ji}}
\affiliation{Dapartment of Physics, Tsinghua University, Beijing, 100084, China}

\author{\mbox{Pavel Fadeev}}
\affiliation{Helmholtz-Institut, GSI Helmholtzzentrum f{\"u}r Schwerionenforschung, Mainz 55128, Germany}
\affiliation{Johannes Gutenberg University, Mainz 55128, Germany}

\author{Dongdong Hu}
\affiliation{
State Key Laboratory of Particle Detection and Electronics, University of Science and Technology of China, Hefei, Anhui 230026, China}

\author{Xinhua Peng}
\email[]{xhpeng@ustc.edu.cn}
\affiliation{
Hefei National Laboratory for Physical Sciences at the Microscale and Department of Modern Physics, University of Science and Technology of China, Hefei, Anhui 230026, China}
\affiliation{
CAS Key Laboratory of Microscale Magnetic Resonance, University of Science and Technology of China, Hefei, Anhui 230026, China}
\affiliation{
Synergetic Innovation Center of Quantum Information and Quantum Physics, University of Science and Technology of China, Hefei, Anhui 230026, China}

\author{Dmitry Budker}
\affiliation{Helmholtz-Institut, GSI Helmholtzzentrum f{\"u}r Schwerionenforschung, Mainz 55128, Germany}
\affiliation{Johannes Gutenberg University, Mainz 55128, Germany}
\affiliation{Department of Physics, University of California, Berkeley, CA 94720-7300, USA}

\maketitle


\section{The response of spin-based amplifier to the exotic spin- and velocity-dependent interactions}
This section presents the analysis of the response of our spin-based amplifier to exotic spin- and velocity-dependent interactions $V_{4+5}$ and $V_{12+13}$.
The principle of spin-based amplifiers is described in Ref.~\cite{jiang2021search}.
Exotic spin- and velocity-dependent interactions are predicted to be produced by moving unpolarized nucleons,
such as a rotating bismuth germanate insulator [$\textrm{Bi}_4 \textrm{Ge}_3 \textrm{O}_{12}$ (BGO)] crystal. 
The exotic interactions induce energy shift of $^{129}$Xe spins in the vapor cell
\begin{equation}
    - \bm{\mu}_{\textrm{Xe}} \cdot \textbf{B}_{j}^{\textrm{exo}} =V_{j},
\end{equation} 
where $\bm{\mu}_{\textrm{Xe}}$ is the magnetic moment of $^{129}$Xe spins, $V_{j}$ represents the potential that we measure ($V_{4+5}$ or $V_{12+13}$) and $\textbf{B}_{j}^{\textrm{exo}}$ is the exotic pseudo-magnetic field produced by the interaction $V_{j}$ (see Sec.~\ref{sec2} for details).

The spin-based amplifier is used to search for such pseudo-magnetic fields.
For simplicity, only a single-frequency component, for example, $\textbf{B}_{\textrm{ac}}^{\textrm{exo}}=B_{\textrm{ac}}^{\textrm{exo}} \cos(2 \pi \nu t)\hat{y}$ of the pseudo-magnetic field $\textbf{B}_j^{\textrm{exo}}$ is considered. 
We first derive the resonant response of the spin-based amplifier to the pseudo-magnetic field $\textbf{B}_{\textrm{ac}}^{\textrm{exo}}$. 
The response is described by the coupled Bloch equations~\cite{jiang2021search}
\begin{align}
    \label{E1}
    &\frac{\partial \textbf{P}^{e}}{\partial t}=\frac{\gamma_{e}}{Q} (B^{0}_z\hat{z}+\beta {M}^{n}_0\textbf{P}^{n})\times\textbf{P}^{e} + \frac{P^{e}_{0} \hat{z} -\textbf{P}^{e} }{T_e Q},\\
    \label{E2}
    &\frac{\partial \textbf{P}^{n}}{\partial t}=\gamma_{n}({B}^{0}_z\hat{z}+\textbf{B}_{\textrm{ac}}^{\textrm{exo}}+\beta {M}^{e}_0\textbf{P}^{e}) \times \textbf{P}^{n} + \frac{P^{n}_{0}\hat{z} -\textbf{P}^{n} }{\{T_{2n},T_{2n},T_{1n}\}},
\end{align} 
where $\textbf{P}^{e}$ ($\textbf{P}^{n}$) is the polarization of $^{87}$Rb electron ($^{129}$Xe nucleus); $\gamma_e$ ($\gamma_n$) is the gyromagnetic radio of the $^{87}$Rb  electron ($^{129}$Xe nucleus); $Q$ is the electron slowing-down factor originated from hyperfine interaction and spin-exchange collisions; $B^0_z\hat{z}$ is the applied bias field; $M^{e}_0$ ($M^{n}_0$) is the maximum magnetization of $^{87}$Rb electron ($^{129}$Xe nucleus) associated with full spin polarizations; $P^e_0$ ($P^n_0$) is the equilibrium polarization of the $^{87}$Rb electron ($^{129}$Xe nucleus); $T_e$ is the common relaxation time of $^{87}$Rb electron spins; and $T_{1n}$ ($T_{2n}$) is the longitudinal (transverse) relaxation time of $^{129}$Xe spins. The Fermi-contact interaction between $^{87}$Rb and $^{129}$Xe spins introduces an effective magnetic field~\cite{walker1997spin,jiang2021floquet,jiang2021search}
\begin{equation}
\textbf{B}^{e,n}_{\textrm{eff}}=\beta M_0^{e,n}\textbf{P}^{e,n},    
\end{equation}
where $\beta = 8\pi \kappa_0 /3$.
Here $\textbf{B}^{e}_{\textrm{eff}}$ ($\textbf{B}^{n}_{\textrm{eff}}$) represents the effective magnetic field experienced by $^{129}$Xe ($^{87}$Rb) spins.

To obtain $\textbf{B}^{e,n}_{\textrm{eff}}$,
we first simplify Eqs.~(\ref{E1}) and (\ref{E2}). 
In the operation of the spin-based amplifier, $^{87}$Rb and $^{129}$Xe spins are under a bias field $B_{z}^{0}\hat{z}$.
The strength of $B_{z}^{0}$ is on the order of 400~nT, which is larger than the strength of $\textbf{B}^{e,n}_{\textrm{eff}}$.
In this situation, $^{87}$Rb and $^{129}$Xe spins are weakly coupled to each other, and Eqs.~(\ref{E1}) and (\ref{E2}) can be greatly simplified. 
The details are explained as follows. 
(1) $P^{e,n}_{z}$ can be approximated as a constant, which is much larger than $P^{e,n}_{x}$ and $P^{e,n}_{y}$. 
Because the applied $B^{0}_{z}$ is much larger than the effective field $\lambda M^{e}_0 {P}_z^{e}$ (on the order of nT), we can further neglect $\lambda M^{e}_0 {P}_z^{e}$ in Eq.~(\ref{E2}).
Therefore, based on Eq.~(\ref{E2}), the $^{129}$Xe spins evolve under the bias field $B^{0}_{z}\hat{z}$. 
As a result, Eq.~(\ref{E2}) can be independently solved at first and $\textbf{P}^n$ can be evaluated. 
(2) Then, we can take the solution of $\textbf{P}^n$ into Eq.~(\ref{E1}) and solve for $\textbf{P}^e$.

We write the total magnetic field experienced by $^{87}$Rb spins as $\textbf{B}={B}^{0}_z\hat{z}+\beta {M}^n_0  \textbf{P}^{n}$ and obtain the simplified Bloch equations,
\begin{eqnarray}
\label{E3}
\frac{\partial \textbf{P}^{e}}{\partial t}&=&\frac{\gamma_{e}}{Q} \textbf{B} \times\textbf{P}^{e} + \frac{P^{e}_{0} \hat{z} -\textbf{P}^{e} }{T_{e} Q},\\
\label{E4}
\frac{\partial \textbf{P}^{n}}{\partial t}&=&\gamma_{n}({B}^{0}_z\hat{z}+\textbf{B}_{\textrm{ac}}^{\textrm{exo}})\times \textbf{P}^{n} +  \frac{P^{n}_{0}\hat{z}-\textbf{P}^{n} }{\{T_{2n},T_{2n},T_{1n}\}}.
\end{eqnarray}
We first solve for the evolution of $^{129}$Xe spins under a bias field $B^{0}_{z} \hat{z}$. 
The Larmor frequency of $^{129}$Xe spins is $\nu_{0}=\gamma_{n}B_{z}^{0}/(2\pi)$. 
With rotating-wave approximation, we can rewrite the Bloch equation of $^{129}$Xe spins in the rotating frame,
\begin{equation}
\begin{aligned}
\dfrac{\partial}{\partial t}\tilde{\textbf{P}}^{n}=\gamma_{n} \tilde{\textbf{B}}^{0} \times \tilde{\textbf{P}}^{n}-\dfrac{\tilde{P}_{x}^{n}\hat{x}+\tilde{P}_{y}^{n}\hat{y}}{T_{2n}}-\dfrac{(P_{z}^{n}-P_{0}^{n})\hat{z}}{T_{1n}},
\label{E5}
\end{aligned}
\end{equation}
where $\tilde{\textbf{B}}^0=\dfrac{2\pi(\nu_{0}-\nu)}{\gamma_{n}}\hat{z}+\dfrac{B_{\textrm{ac}}^{\textrm{exo}}}{2}\hat{y}$ is the equivalent magnetic field in the rotating frame. We derive the steady-state solution of three components of $\tilde{\textbf{P}}^{n}$ by solving Eq.~(\ref{E5}) in rotating frame and transforming back to the laboratory frame,
\begin{align}
\label{E6}
&P^{n}_{x}=\dfrac{1}{2}P^{n}_{0} \gamma_{n} B_{\textrm{ac}}^{\textrm{exo}}\dfrac{ T_{2n}\cos(2\pi\nu t)+2\pi(\nu-\nu_{0})T_{2n}^{2}\sin(2\pi\nu t)}{1+({\gamma_{n} B_{\textrm{ac}}^{\textrm{exo}}}/2)^{2}T_{1n}T_{2n}+[2\pi(\nu-\nu_{0})]^{2}T_{2n}^{2}},\\
\label{E7}
&P^{n}_{y}=\dfrac{1}{2}P^{n}_{0} \gamma_{n} B_{\textrm{ac}}^{\textrm{exo}}\dfrac{T_{2n}\sin(2\pi\nu t)-2\pi(\nu-\nu_{0})T_{2n}^{2}\cos(2\pi\nu t)}{1+({\gamma_{n} B_{\textrm{ac}}^{\textrm{exo}}}/2)^{2}T_{1n}T_{2n}+[2\pi(\nu-\nu_{0})]^{2}T_{2n}^{2}},\\
\label{E8}
&P^n_z=P^n_0 \frac{1+[2\pi(\nu-\nu_0)T_{2n}]^2}{1+({\gamma_{n} B_{\textrm{ac}}^{\textrm{exo}}}/2)^{2}T_{1n}T_{2n}+[2\pi(\nu-\nu_{0})T_{2n}]^{2}},
\end{align}
where $C$ is a constant determined by initial conditions. 
According to $\textbf{B}_{\rm{eff}}^n=\beta M^{n}_0 \textbf{P}^{n}$, we can derive the effective field experienced by $^{87}$Rb spins as
\begin{equation}
\begin{aligned}
\label{E9}
&\textbf{B}_{\rm{\rm{eff}}}^n=\overbrace{\dfrac{1}{2}\beta M^{n}_0 P^{n}_{0} \gamma_{n} B_{\textrm{ac}}^{\textrm{exo}}\dfrac{T_{2n}\cos(2\pi\nu t)+2\pi(\nu-\nu_{0})T_{2n}^{2}\sin(2\pi\nu t)}{1+({\gamma_{n} B_{\textrm{ac}}^{\textrm{exo}}}/2)^{2}T_{1n}T_{2n}+[2\pi(\nu-\nu_{0})]^{2}T_{2n}^{2}}\hat{x}}^\textrm{ effective field generated by $^{129}$Xe $x$ magnetization}
+\overbrace{\dfrac{1}{2}\beta M^{n}_0 P^{n}_{0} \gamma_{n} B_{\textrm{ac}}^{\textrm{exo}}\dfrac{T_{2n}\sin(2\pi\nu t)-2\pi(\nu-\nu_{0})T_{2n}^{2}\cos(2\pi\nu t)}{1+({\gamma_{n} B_{\textrm{ac}}^{\textrm{exo}}}/2)^{2}T_{1n}T_{2n}+[2\pi(\nu-\nu_{0})]^{2}T_{2n}^{2}}\hat{y}}^\textrm{effective field generated by $^{129}$Xe $y$ magnetization}.
\end{aligned}
\end{equation}
Such an effective field can be measured $in$ $situ$ by the $^{87}$Rb magnetometer~\cite{jiang2021search,jiang2021floquet}.
In the following, we derive the amplification factor and bandwidth of the pseudo-magnetic field generated by the spin- and velocity-dependent interactions with a spin-based amplifier.

~\

\noindent
\textbf{Amplification factor}.~As seen in Eq.~(\ref{E9}), the pseudo-magnetic field can induce an oscillating $^{129}$Xe nuclear magnetization, generating a considerable effective magnetic field $\textbf{B}_{\rm{eff}}^n$ on $^{87}$Rb spins.
Moreover, the signal from $\textbf{B}_{\rm{eff}}^n$ can be much larger than that from the oscillating pseudo-magnetic field $B^{\rm{exo}}_{\rm{ac}}$.
We define an amplification factor:
\begin{equation}
\eta=|\textbf{B}_{\rm{eff}}^n|/|\textbf{B}^{\textrm{exo}}_{\rm{ac}}|.
\label{E12}
\end{equation}
To derive the amplification factor $\eta$ on resonance, we assume that the pseudo-magnetic field strength is small and thus the term $({\gamma_{n} B_{\textrm{ac}}^{\textrm{exo}}}/2)^{2} T_{1n} T_{2n}$ can be neglected in Eq.~(\ref{E9}). 
In this situation, $\textbf{B}_{\rm{eff}}^n$ can be written as
\begin{equation}
	 \textbf{B}_{\textrm{eff}}^n(\nu=\nu_0)=\dfrac{4\pi}{3}\kappa_0 M^{n}_0 P^{n}_{0} \gamma_{n} T_{2n}[\cos(2\pi\nu t)\hat{x} + \sin(2\pi\nu t)\hat{y}]B_{\textrm{ac}}^{\textrm{exo}}.
	 \label{beff}
\end{equation}
Based on Eq.~(\ref{beff}), the effective field $\textbf{B}_{\textrm{eff}}^n$ is a circularly polarized field and its amplitude is equal to $\dfrac{4\pi}{3}\kappa_0 M^{n}_0 P^{n}_{0} \gamma_{n} T_{2n} \cdot B_{\textrm{ac}}^{\textrm{exo}}$.
Lastly, the amplification factor is
\begin{equation}
	\eta=\dfrac{4\pi}{3}\kappa_0 M^{n}_0 P^{n}_{0} \gamma_{n} T_{2n}.
	\label{eta}
\end{equation}
As described in the main text [see Fig.~2(a)], $\eta$ is calibrated as 116 in our $^{129}$Xe spin-based amplifier.
This indicates that the corresponding sensitivity to the signal from the spin- and velocity-dependent interactions can be improved by two orders of magnitude. 
In the near future, further improvement of the search sensitivity can be anticipated using a $^3$He-K system.  
Based on theoretical calculations~\cite{jiang2021search}, the amplification factor could be as large as 10$^4$ in a $^3$He-K system. 
The sensitivity to the signal from $\textbf{B}_{\textrm{ac}}^{\textrm{exo}}$ can be improved by four orders of magnitude and can potentially reach a few aT/$\sqrt{\rm{Hz}}$.
Accordingly, it is possible to establish more stringent constraints on $f_{4+5}$ and $f_{12+13}$ through the $^3$He-K system.

~\

\noindent
\textbf{Bandwidth}.~Up to now, we have only considered the amplification factor in the resonant case. 
Next, we analyse the amplification performance in the near-resonant case.
Based on Eq.~(\ref{E9}), the amplification factor reaches maximum on resonance, and rapidly decreases when the oscillation frequency of applied magnetic field $\nu$ is far off-resonance. 
Therefore, there is a limited bandwidth for our spin-based amplifier. 
Based on Eq.~(\ref{E9}), we can rewrite the equation in the amplitude spectrum,
\begin{equation}
\begin{aligned}
|\textbf{B}_{\textrm{eff}}^n(\nu)| \propto \dfrac{\Lambda/2}{\sqrt{(\nu-\nu_{0})^2+(\Lambda/2)^2}}.
\label{E15}
\end{aligned}
\end{equation}
We assume that the pseudo-magnetic field amplitude is weak and thus the term $({\gamma_{n} B_{\textrm{ac}}^{\textrm{exo}}}/2)^{2}T_{1n}T_{2n}$ can be neglected in Eq.~(\ref{E9}). 
As a demonstration, we set the bias field as $B_{z}^{0} \approx 423$~nT and the simulated oscillating field strength $\approx 13$~pT. 
The full width at half maximum (FWHM) is $\approx$~13~mHz [see Fig.~2(a) in main text].
Thus, the spin-based amplifier can enhance the signal in a correspondingly narrow frequency range, which is well suited for resonantly searching for exotic spin- and velocity-dependent interactions.
To precisely adjust and steadily maintain the operation frequency $\nu_0 \approx \nu$ of the spin-based amplifier, we use a precision current source (Krohn-Hite Model 523) to provide the bias field $B_z^0 \hat{z}$.
Moreover, the vapor cell is magnetically shielded with a five-layer cylindrical $\mu$-metal shield (shielding factor of $10^6$) to avoid fluctuations of the operation frequency from the ambient magnetic field.

\section{Numerical simulations of the exotic pseudo-magnetic fields}
\label{sec2}
This section presents numerical simulations of the pseudo-magnetic fields generated by the exotic spin- and velocity-dependent interactions.
The exotic interactions studied here are~\cite{dobrescu2006spin,leslie2014prospects}
\begin{flalign}
    \label{E16}
    &V_{4+5}=-f_{4+5}\frac{\hbar^2}{8 \pi m c}[\hat{\sigma} \cdot (\bm{v} \times \hat{r})]\left(\frac{1}{\lambda r}+\frac{1}{r^2}\right)e^{-r/\lambda}, \\
    \label{E17}
    &V_{12+13}=f_{12+13}\frac{\hbar}{8 \pi}(\hat{\sigma} \cdot \bm{v})\left (\frac{1}{r}\right )e^{-r/\lambda},
\end{flalign}
where $f_{4+5}$, $f_{12+13}$ are dimensionless coupling constant, $c$ is the speed of light in vacuum, $\hat{\sigma}$ is the spin vector and $m$ is the mass of the polarized fermion, $\bm{v}$ is the relative velocity between two interacting fermions, $\hat{r}$ is the unit vector in the direction between them, and $\lambda \!= \!\hbar(m_b c)^{-1}$ is the force range (or the boson Compton wavelength) with $m_b$ being the light boson mass.

\begin{figure} [h] 
	\makeatletter
	\def\@captype{figure}
	\makeatother
	\includegraphics[scale=1.2]{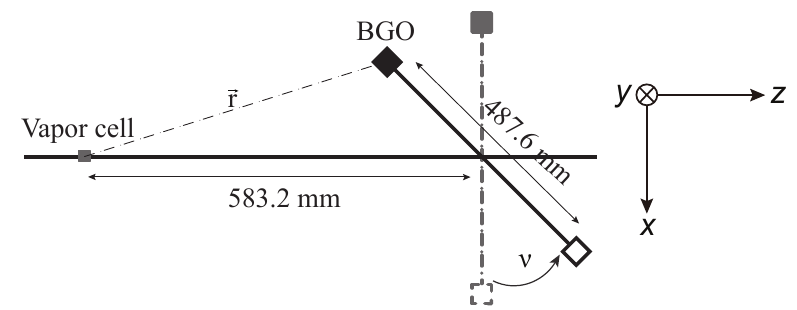}
	\caption{\textbf{Experimental setup}. The exotic spin- and velocity-dependent interactions are generated by the single rotating BGO [bismuth germanate insulator $\textrm{Bi}_4 \textrm{Ge}_3 \textrm{O}_{12}$] crystal. The single BGO crystal is connected to one end of a symmetrical aluminum rod. The rotation frequency is $\nu \approx 4.997$~Hz. The detailed information for the setup is presented in the main text.}
	\label{setup}
\end{figure}

Here we consider the situation where the pseudo-magnetic field is generated by a single rotating BGO crystal, as shown in Fig.~\ref{setup}.
Driven by a servo motor, the 112.34-g BGO crystal containing $6.71 \times 10^{25}$ nucleons connected to a 48.76-cm aluminum rod rotates with frequency $\nu \approx 4.997$~Hz in the $xz$ plane. 
The center of the aluminum rod is located 58.32~cm away from the center of the $^{129}$Xe vapor cell.
In the following, we numerically simulate the exotic pseudo-magnetic fields $\textbf{B}^{\textrm{exo}}_{4+5}$ and $\textbf{B}^{\textrm{exo}}_{12+13}$.

\begin{figure} [t] 
	\makeatletter
	\def\@captype{figure}
	\makeatother
	\includegraphics[scale=1.05]{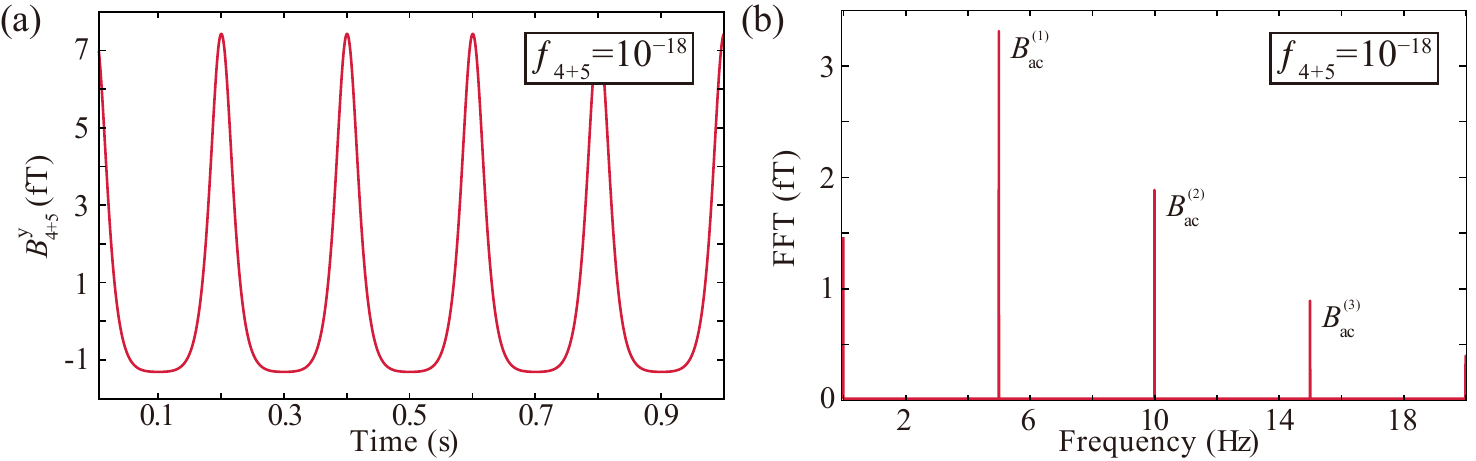}
	\caption{\textbf{Simulations of $V_{4+5}$ produced by a BGO crystal}. Signal from the simulated pseudo-magnetic field $\textbf{B}_{4+5}^{\textrm{exo}}$ is generated by the rotating BGO [bismuth germanate insulator $\textrm{Bi}_4 \textrm{Ge}_3 \textrm{O}_{12}$] crystal with the assumption of $\lambda=1.0$~m and $f_{4+5}=10^{-18}$. (a) Time-domain signal of $B_{4+5}^y$. (b) Fourier transformation spectrum of $B_{4+5}^y$.}
	\label{V4}
\end{figure}

\begin{figure} [b] 
	\makeatletter
	\def\@captype{figure}
	\makeatother
	\includegraphics[scale=1.05]{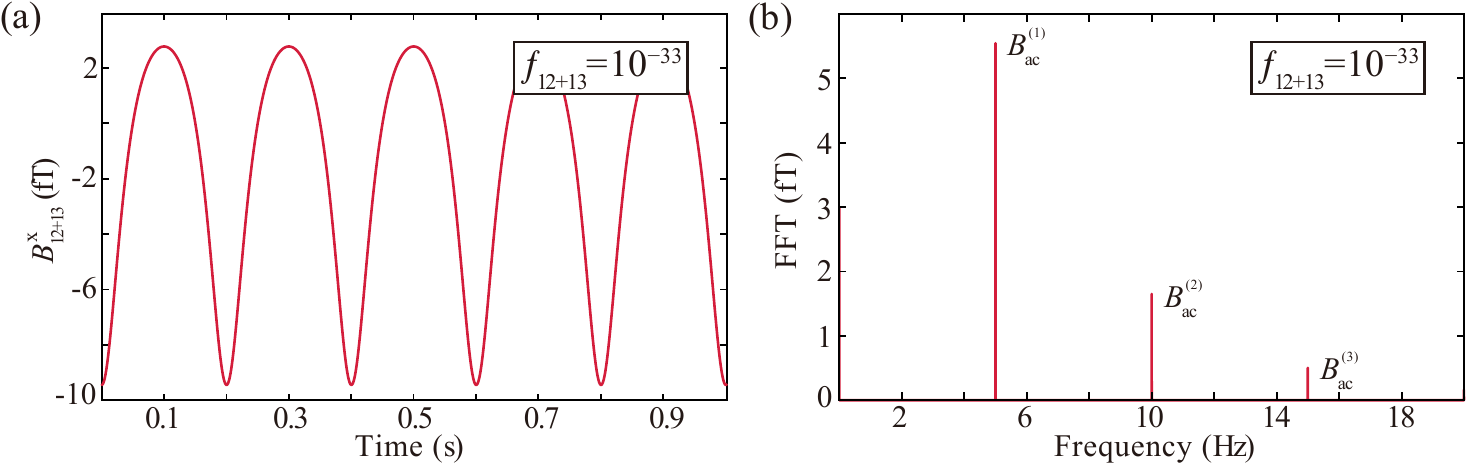}
	\caption{\textbf{Simulations of $V_{12+13}$ produced by a BGO crystal}. Signal from the pseudo-magnetic field $\textbf{B}_{12+13}^{\textrm{exo}}$ is generated by the rotating BGO [bismuth germanate insulator $\textrm{Bi}_4 \textrm{Ge}_3 \textrm{O}_{12}$] crystal with the assumption of $\lambda=1.0$~m and $f_{12+13}=10^{-33}$. (a) Time-domain signal of $B_{12+13}^x$. (b) Fourier transformation spectrum of $B_{12+13}^x$.}
	\label{V12}
\end{figure}

~\

\noindent
\textbf{Simulation of} \bm{$V_{4+5}$}.~Here we consider the pseudo-magnetic field $\textbf{B}_{4+5}^{\textrm{exo}}$ generated by a single rotating BGO crystal.
The pseudo-magnetic field can be obtained by integrating over the volume of the single BGO crystal ($25.0\times25.0\times25.0$~mm$^3$),
\begin{equation}
\begin{split}
     \textbf{B}_{4+5}^{\textrm{exo}}=f_{4+5} \frac{\hbar^2}{8 \pi m c |\bm{\mu}_{\textrm{Xe}}|} \iiint \rho(\bm{r})(\bm{v} \times \hat{r})\left(\frac{1}{\lambda r}+\frac{1}{r^2}\right)e^{-r/\lambda}d \bm{r},
\label{E18}
\end{split}
\end{equation}
where $\rho(\bm{r})$ is the BGO crystal's nucleon density at location $\bm{r}$ and $m$ is the mass of a neutron.
By numerical simulation with the assumption of $f_{4+5}=10^{-18}$ and $\lambda=1.0$~m, we show that the pseudo-magnetic field $\textbf{B}_{4+5}^{\textrm{exo}}=B^y_{4+5}\hat{y}$ is not a pure trigonometric function. 
Figure~\ref{V4} shows that $B^y_{4+5}\hat{y}$ contains harmonic frequencies (see main text) at $\nu,2\nu,3\nu,$....
The ratios of the field strengths at harmonic frequencies are $\!B_{\textrm{ac}}^{(1)}\!:\!B_{\textrm{ac}}^{(2)}\!:\!B_{\textrm{ac}}^{(3)}\! \approx \!5.1\!:\!2.9\!:1.4\!$, as shown in Fig.~\ref{V4}(b).
Due to the relatively narrow bandwidth of the spin-based amplifier, only a single-frequency component can be resonantly amplified.
Accordingly, we choose the dominant ($N=1$) first harmonic $B_{\textrm{ac}}^{(1)} \cos(2 \pi \nu t)\hat{y}$ at $\nu$ to be measured and filter out other harmonics.


~\

\noindent
\textbf{Simulation of} \bm{$V_{12+13}$}.~The pseudo-magnetic field of $V_{12+13}$ can be derived by the same method demonstrated above, 
\begin{equation}
      \textbf{B}_{12+13}^{\textrm{exo}}=-f_{12+13} \frac{\hbar}{8 \pi  |\bm{\mu}_{\textrm{Xe}}|} \iiint \rho(\hat{r})(\bm{v} )\left(\frac{1}{r}\right)e^{-r/\lambda}d\bm{r}.
\end{equation}
In contrast to $\textbf{B}_{4+5}^{\textrm{exo}}$, the pseudo-magnetic field $\textbf{B}_{12+13}^{\textrm{exo}}=B_{12+13}^x \hat{x}+B_{12+13}^z \hat{z}$ contains two components along $x$ and $z$.
We note that the spin-based amplifier is insensitive to the oscillating field along $z$, because the oscillating field along $z$ can not induce $^{129}$Xe transverse magnetization. 
Therefore, only $x$ component of the pseudo-magnetic field $B_{12+13}^x\hat{x}$ should be considered. 
Based on our numerical simulation with the assumption of $\lambda=1.0$~m and $f_{12+13}=10^{-33}$,
the ratios of the field strengths at harmonic frequencies are  $\!B_{\textrm{ac}}^{(1)}\!:\!B_{\textrm{ac}}^{(2)}\!:\!B_{\textrm{ac}}^{(3)}\! \approx \!5.5\!:\!1.7\!:0.5\!$, as shown in Fig.~\ref{V4}(b).
In the same way, we choose the dominant ($N=1$) first harmonic $B_{\textrm{ac}}^{(1)} \cos(2 \pi \nu t)\hat{x}$ at $\nu$ to be measured and filter out other harmonics, as shown in Fig.~\ref{V12}(b).

~\

\begin{figure} [h] 
	\makeatletter
	\def\@captype{figure}
	\makeatother
	\includegraphics[scale=1.05]{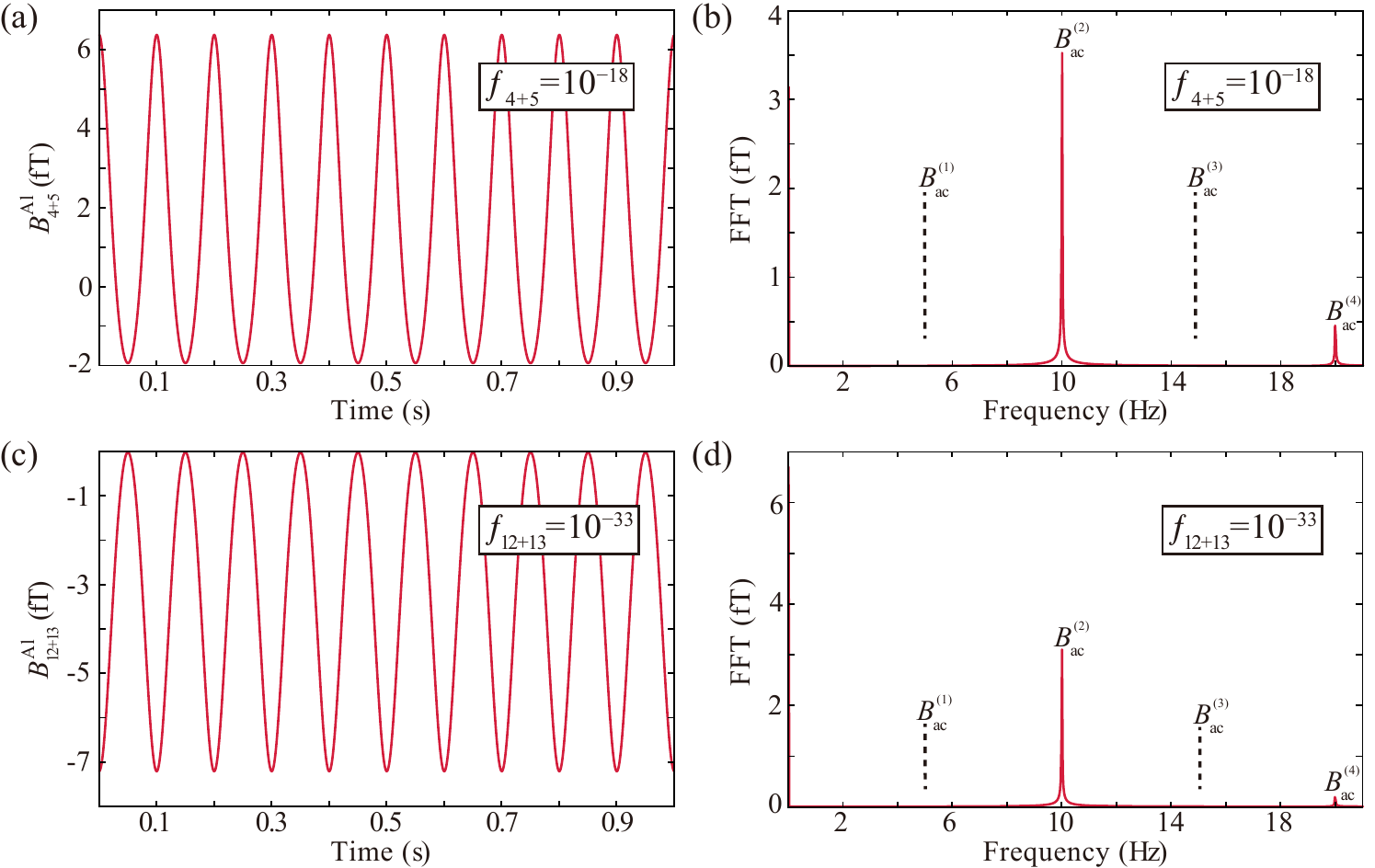}
	\caption{\textbf{Simulations of $V_{4+5}$ and $V_{12+13}$ produced by an aluminum rod}. Signal of the pseudo-magnetic field is generated by the rotating aluminum rod with the assumption of $\lambda=1.0$~m and $f_{4+5}=10^{-18}$ or $f_{12+13}=10^{-33}$. (a) Time-domain signal of $B_{4+5}^{\textrm{Al}}$. (b) Fourier transformation spectrum of $B_{4+5}^{\textrm{Al}}$. (c) Time-domain signal of $B_{12+13}^{\textrm{Al}}$. (d) Fourier transformation spectrum of $B_{12+13}^{\textrm{Al}}$.  }
	\label{al}
\end{figure}

\noindent
\textbf{Simulation of the exotic interactions generated by the aluminum rod}.~In above discussion, we do not consider the spin- and velocity-dependent interactions generated from another unpolarized mass, the rotating aluminum rod.
The 610.34-g aluminum rod ($487.6\times30.5\times15.2$~mm$^3$) contains $3.64 \times 10^{26}$ unpolarized nucleons, which can also induce the energy shift of $^{129}$Xe spins.
Therefore, we now analyze the pseudo-magnetic field generated by the rotating aluminum rod. 
Figure~\ref{al} shows the Fourier transformation spectrum of the oscillating pseudo-magnetic field $B^y_{4+5}\hat{y}$ and $B^x_{12+13}\hat{x}$ by our numerical simulation with the assumption of $\lambda=1.0$~m and $f_{4+5}=10^{-18}$ or $f_{12+13}=10^{-33}$.
It is worth noting that the pseudo-magnetic fields produced by the aluminum rod do not contain odd harmonics at  $\nu,3\nu,5\nu,...$ and only contain even harmonics at $2\nu,4\nu,6\nu,...$.
The reason is that the configuration of the rotating aluminum is central-symmetrical around the rotation axis and axial-symmetrical along $z$, as shown in Fig.~\ref{setup}.
In one period, one end of the aluminum rod comes to the nearest position to the vapor cell twice, the pseudo-magnetic field reaches maximum twice, and the dominant frequency becomes $2\nu$.

In our experiment, we search for the pseudo-magnetic field at $\nu$ generated by the single BGO crystal and do not consider the ones at $2\nu$ generated by the aluminum rod.
First, as discussed in main text, to avoid the air vibration at $2\nu$, we choose to measure the signal at $\nu$.
Therefore, the operation frequency of the spin-based amplifier is set at $\nu$.
Second, due to the relatively narrow bandwidth of the spin-based amplifier, it only amplifies the signal at $\nu$ and the signal at $2\nu$ is negligible.
Therefore, we do not need to consider the pseudo-magnetic field generated by the aluminum rod.

\bibliographystyle{naturemag}
\bibliography{supplementrefs.bib}